# Two components of the macroscopic general field

Sergey G. Fedosin

PO box 614088, Sviazeva str. 22-79, Perm, Perm Krai, Russia

E-mail: intelli@list.ru

The general field, containing all the macroscopic fields in it, is divided into the mass component, the source of which is the mass four-current, and the charge component, the source of which is the charge four-current. The mass component includes the gravitational field, acceleration field, pressure field, dissipation field, strong interaction and weak interaction fields, other vector fields. The charge component of the general field represents the electromagnetic field. With the help of the principle of least action we derived the field equations, the equation of the matter's motion in the general field, the equation for the metric, the energy and momentum of the system of matter and its fields, and calibrated the cosmological constant. The general field components are related to the corresponding vacuum field components so that the vacuum field generates the general field at the macroscopic level.
***Keywords:*** *general field; vacuum field; acceleration field; pressure field; dissipation field.*

## 1. Introduction

Most of the unified field theories, such as the theory of everything, grand unified theory, loop quantum gravity, string theory, and some other theories, are based on the quantum approach and are intended to unite the fundamental interactions at the level of elementary particles. There were known attempts to unite the physical fields at the macroscopic level for sufficiently massive bodies, where gravitation becomes the defining interaction. A variety of approaches are used for this purpose. Thus, in the Kaluza-Klein theory [1-2] the so far undiscovered fifth dimension and some scalar field were added to the ordinary four-dimensional spacetime in order to derive both the gravitational field equations and the equations that are equivalent to the Maxwell equations. The unified field theories, in four-dimensions and extra spacial dimensions, were considered in several important papers [3-6].

In [7] it is assumed that the field strength vector and the solenoidal vector of the unified field consist of the sum of the field strengths and solenoidal vectors of all the known fields with the corresponding coefficients. In this case the unified field obeys the Maxwell equations, in which the source of the unified field strength is the sum of the products of the fields' charge densities by certain coefficients, and the source of the unified field's solenoidal vector is the sum of the products of the currents with certain coefficients. As a result, each four-current, associated with a particular field, contributes to the unified field vector components. The drawback of this approach is lack of covariance of the presented unified field equations.

The concept of the general force vector field, which includes the electromagnetic and gravitational fields, acceleration field, pressure field, dissipation field, strong interaction field, weak interaction field, and other vector fields, was presented in [8]. The characteristic feature of this approach is that the four-potential of the general field is given by the sum of the four-potentials of all particular fields. An exception is the four-potential of the electromagnetic field, which is included in the sum of the four-potentials with a coefficient equal to the charge density to mass density ratio $\rho_{0q}/\rho_0$. The equations derived from the principle of least action fully describe the general field and its interaction with the matter.

A certain limitation of this approach is that the density ratio $\rho_{0q}/\rho_0$ in the physical system under consideration is assumed to be unchanged. Below, we present a more complete and universal theory, in which the general field is resolved into two main components. The source of the first component is the mass four-current $J^\mu$, which generates such vector fields as the gravitational field, acceleration field, pressure field, dissipation field, macroscopic fields of strong and weak interactions. The second component of the general field is the electromagnetic field, the source of which is the charge four-current $j^\mu$.

A characteristic feature of macroscopic fields is that their description may not coincide with the field's description at the microscopic level. Thus, the general theory of relativity (GTR) is considered to be quite satisfactory, although it is not related to the quantum theory of gravity, and it does not follow from it. From the set of axioms of GTR it follows that the gravitational field is described by the metric tensor and the stress-energy pseudotensor, which are determined based on the space-time geometry [9]. This approach, which was many times confirmed by experiments, leads to the idea that the gravitational field is a tensor (metric) field. Thus, the gravitation in GTR does not reduce to the vector field, and the tensor character of the field directly preconditions the absence of the dipole component in the gravitational radiation.

In order to describe the gravitation we use the axiomatically constructed covariant theory of gravitation (CTG) with the vector four-potential, gravitational tensor and stress-energy tensor [10]. In this case the gravitation is an independent physical field that does not require full reduction to geometry. In CTG for each individual body the dipole gravitational radiation is admitted. Since for detection of gravitational effects at least two bodies are required, their dipole radiation mutually cancels out, and the system of bodies' radiation always occurs in a quadrupole way or by higher multipoles. Thus, with respect to radiation the tensor field in GTR does not contradict the vector field in CTG. The difference between the basic postulates of the both theories does not prevent from describing in CTG all the standard effects of GTR [10], presenting the expression for the metric tensor [11], explaining the Pioneer anomaly [12], and considering for macroscopic systems the contribution of the gravitational field in the Navier-Stokes equation [13], in estimation of the parameters of cosmic bodies [14] and in the virial theorem [15].

Although strong and weak interactions at the level of elementary particles are usually described by vector non-Abelian fields (nonlinear Yang-Mills fields with self-action), at the macroscopic level we believe it is possible to describe these interactions with the help of ordinary four-dimensional vector fields. Indeed, quantum effects at the macrolevel as a rule disappear due to the large number of interacting particles, as well as the contributions of individual particles into the fields, only the average values of fields become important. As a result, in order to describe the macroscopic fields it suffices to use simple schemes in the form of equations for the four-potentials of Maxwellian type, at least as a first approximation.

## 2. The action function and its variation

We assume that the two components of the general field and the corresponding four-currents are sufficiently independent of each other. This allows us to apply the superposition principle in the principle of least action, so that each component independently contributes to the Lagrangian. The action function for the continuously distributed matter located in the general field (both in the proper field and the external field) in the curved spacetime similarly as in [8], [16] takes the following form:

$$S = \int L\, dt = \int \left( k(R - 2\Lambda) - \frac{1}{c} s_\mu J^\mu - \frac{c}{16\pi\varpi} s_{\mu\nu} s^{\mu\nu} - \frac{1}{c} A_\mu j^\mu - \frac{c\varepsilon_0}{4} F_{\mu\nu} F^{\mu\nu} \right) \sqrt{-g}\, d\Sigma,$$

(1)

where $L$ is the Lagrangian, $R$ is the scalar curvature, $\Lambda$ is the cosmological constant, $J^\mu = \rho_0 u^\mu$ is the four-vector of the mass current, $\rho_0$ is the mass density of a point particle of matter in the reference frame associated with the particle, $u^\mu = \dfrac{c\, dx^\mu}{ds}$ is the four-velocity of the point particle, $c$ is the speed of light, $s_\mu = \left( \dfrac{\theta}{c}, -\mathbf{\Phi} \right)$ is the four-potential of the mass component of the general field, described with the help of the scalar potential $\theta$ and the vector potential $\mathbf{\Phi}$ of this field, $s_{\mu\nu}$ is the tensor of the mass component of the general field, $A_\mu = \left( \dfrac{\varphi}{c}, -\mathbf{A} \right)$ is the four-potential of the electromagnetic field defined using the scalar potential $\varphi$ and the vector potential $\mathbf{A}$, $j^\mu = \rho_{0q} u^\mu$ is the four-vector of the charge current, $\rho_{0q}$ is the charge density of a point particle in the reference frame associated with the particle, $\varepsilon_0$ is the vacuum permittivity, $F_{\mu\nu}$ is the electromagnetic tensor, $k$ and $\varpi$ are considered the constant coefficients.

In (1) the first term under the integral sign is proportional to the energy density associated with the curvature and the cosmological constant. The second term defines the energy density of the mass four-current at the four-potential $s_\mu$. Similarly, the fourth term for the electromagnetic field defines the energy density of the charge four-current at the four-potential $A_\mu$. The third and fifth terms are associated with the energy density of the general field's mass components and the energy density of the electromagnetic field, respectively, and they do not vanish even in the empty space outside the matter.

The four-potential of the mass component of the general field is defined as a generalized four-velocity in the form of the sum of the four-potentials of the gravitational field [17], acceleration field and pressure field [18], dissipation field [13] and fields of strong and weak interactions [8], respectively:

$$s_\mu = D_\mu + u_\mu + \pi_\mu + \lambda_\mu + g_\mu + w_\mu. \tag{2}$$

From (2) and the components of the four-potential $s_\mu$ it follows that the scalar $\theta$ and vector $\boldsymbol{\Phi}$ potentials of the mass component of the general field are the sums of the respective scalar and vector potentials of the fields under consideration.

The tensor of the mass component of the general field $s_{\mu\nu}$ is defined as the four-curl of the four-potential $s_\mu$. In view of (2), the tensor $s_{\mu\nu}$ is expressed in terms of the sum of the tensors of the gravitational field, acceleration field, pressure field, dissipation field, and fields of strong and weak interactions, respectively:

$$s_{\mu\nu} = \nabla_\mu s_\nu - \nabla_\nu s_\mu = \partial_\mu s_\nu - \partial_\nu s_\mu =$$
$$= \partial_\mu \left( D_\nu + u_\nu + \pi_\nu + \lambda_\nu + g_\nu + w_\nu \right) - \partial_\nu \left( D_\mu + u_\mu + \pi_\mu + \lambda_\mu + g_\mu + w_\mu \right) = \qquad (3)$$
$$= \Phi_{\mu\nu} + u_{\mu\nu} + f_{\mu\nu} + h_{\mu\nu} + \gamma_{\mu\nu} + w_{\mu\nu}.$$

The action function with the terms similar to the terms in (1) was varied in a number of works, for example, in [16], [19]. For the action function variation we can write the following:

$$\delta S = \delta S_1 + \delta S_2 + \delta S_3 + \delta S_4 + \delta S_5 = 0, \qquad (4)$$

$$\delta S_1 = \int \left( -k R^{\alpha\beta} + \frac{k}{2} R g^{\alpha\beta} - k \Lambda g^{\alpha\beta} \right) \delta g_{\alpha\beta} \sqrt{-g} \, d\Sigma,$$

$$\delta S_2 = \int \left( -\frac{1}{c} s_{\beta\sigma} J^\sigma \xi^\beta - \frac{1}{2c} s_\mu J^\mu g^{\alpha\beta} \delta g_{\alpha\beta} - \frac{1}{c} J^\beta \delta s_\beta \right) \sqrt{-g} \, d\Sigma,$$

$$\delta S_3 = \int \left( \frac{c}{4\pi\varpi} \nabla_\alpha s^{\alpha\beta} \delta s_\beta - \frac{1}{2c} T^{\alpha\beta} \delta g_{\alpha\beta} \right) \sqrt{-g} \, d\Sigma,$$

$$\delta S_4 = \int \left( -\frac{1}{c} F_{\beta\sigma} j^\sigma \xi^\beta - \frac{1}{2c} A_\mu j^\mu g^{\alpha\beta} \delta g_{\alpha\beta} - \frac{1}{c} j^\beta \delta A_\beta \right) \sqrt{-g} \, d\Sigma,$$

$$\delta S_5 = \int \left( c\varepsilon_0 \nabla_\alpha F^{\alpha\beta} \delta A_\beta - \frac{1}{2c} W^{\alpha\beta} \delta g_{\alpha\beta} \right) \sqrt{-g}\, d\Sigma,$$

where $R^{\alpha\beta}$ is the Ricci tensor, $\delta g_{\alpha\beta}$ is the metric tensor variation, $\sqrt{-g}\, d\Sigma = \sqrt{-g}\, c\, dt\, dx^1 dx^2 dx^3$ is the invariant four-volume, expressed in terms of the time coordinate differential $dx^0 = c\, dt$, the product $dx^1 dx^2 dx^3$ of the space coordinate differentials and the square root $\sqrt{-g}$ of the determinant $g$ of the metric tensor, taken with a negative sign,

$\xi^\beta = \delta x^\beta$ represents a variation of four-coordinates [19, 20], due to which we obtain a variations of the mass four-current $\delta J^\mu$ and the charge four-current $\delta j^\mu$, $\delta s_\beta$ is a variation of the four-potential of the mass component of the general field, $\delta A_\beta$ is a variation of the four-potential of the electromagnetic field.

The stress-energy tensor $T^{\alpha\beta}$ of the mass component of the general field and the stress-energy tensor $W^{\alpha\beta}$ of the electromagnetic field are given by:

$$T^{\alpha\beta} = \frac{c^2}{4\pi\varpi}\left( -g^{\alpha\nu} s_{\kappa\nu} s^{\kappa\beta} + \frac{1}{4} g^{\alpha\beta} s_{\mu\nu} s^{\mu\nu} \right).$$

$$W^{\alpha\beta} = \varepsilon_0 c^2 \left( -g^{\alpha\nu} F_{\kappa\nu} F^{\kappa\beta} + \frac{1}{4} g^{\alpha\beta} F_{\mu\nu} F^{\mu\nu} \right).$$

The properties of the stress-energy tensor $T^{\alpha\beta}$ are described in [8].

## 3. The equations for the fields and the motion of matter

Summing up in (4) the terms with the same variations and equating these sums to zero, we obtain the corresponding equations. In particular, we find the equations for the general field's mass component with the field source in the form of the mass four-current, as well as the equations of this field without sources, resulting from antisymmetry of the tensor $s_{\mu\nu}$:

$$\nabla_\beta s^{\alpha\beta} = -\frac{4\pi\varpi}{c^2} J^\alpha, \qquad \varepsilon^{\alpha\beta\gamma\delta} \nabla_\gamma s_{\alpha\beta} = 0. \tag{5}$$

The electromagnetic field equations have the standard form:

$$\nabla_\beta F^{\alpha\beta} = -\frac{1}{c^2 \varepsilon_0} j^\alpha, \qquad \varepsilon^{\alpha\beta\gamma\delta} \nabla_\gamma F_{\alpha\beta} = 0. \tag{6}$$

Applying the covariant derivative $\nabla_\alpha$ to the equations of the field with the sources in (5) and (6) gives continuity equations for the four-currents in the curved spacetime:

$$R_{\mu\alpha} s^{\mu\alpha} = \frac{4\pi\varpi}{c^2} \nabla_\alpha J^\alpha, \qquad R_{\mu\alpha} F^{\mu\alpha} = \frac{1}{c^2 \varepsilon_0} \nabla_\alpha j^\alpha.$$

The gauge condition of the four-potentials:

$$\nabla_\beta s^\beta = \nabla^\mu s_\mu = 0, \qquad \nabla_\beta A^\beta = \nabla^\mu A_\mu = 0.$$

The equations of motion are obtained from (4) by equating the sum of the terms containing the variation $\xi^\beta$ to zero. In view of (3) we have:

$$F_{\beta\sigma} j^\sigma + s_{\beta\sigma} J^\sigma = F_{\beta\sigma} j^\sigma + \Phi_{\beta\sigma} J^\sigma + u_{\beta\sigma} J^\sigma + f_{\beta\sigma} J^\sigma + h_{\beta\sigma} J^\sigma + \gamma_{\beta\sigma} J^\sigma + w_{\beta\sigma} J^\sigma = 0.$$

(7)

The tensor product $u_{\beta\sigma} J^\sigma$ can be expressed in terms of the four-acceleration $a_\beta$ using the operator of the proper time derivative [18]:

$$-u_{\beta\sigma} J^\sigma = -\rho_0 u^\sigma \left( \nabla_\beta u_\sigma - \nabla_\sigma u_\beta \right) = \rho_0 u^\sigma \nabla_\sigma u_\beta = \rho_0 \frac{D u_\beta}{D\tau} = \rho_0 \frac{d u_\beta}{d\tau} - \rho_0 \Gamma^\lambda_{\sigma\beta} u_\lambda u^\sigma = \rho_0 a_\beta.$$

With this in mind, (7) is transformed into the four-dimensional equation of motion of the viscous compressible and charged fluid [13], with the addition from the density of the four-forces, arising from the strong and weak interactions:

$$\rho_0 a_\beta = F_{\beta\sigma} j^\sigma + \Phi_{\beta\sigma} J^\sigma + f_{\beta\sigma} J^\sigma + h_{\beta\sigma} J^\sigma + \gamma_{\beta\sigma} J^\sigma + w_{\beta\sigma} J^\sigma.$$

Because, due to the tensors' properties, for the corresponding four-forces, the following relations hold true:

$$F_{\beta\sigma} j^\sigma = -\nabla^k W_{\beta k}, \qquad s_{\beta\sigma} J^\sigma = -\nabla^k T_{\beta k},$$

then the equation of motion (7) can be written using the divergence of the sum of the stress-energy tensors:

$$F_{\beta\sigma} j^\sigma + s_{\beta\sigma} J^\sigma = -\nabla^k (W_{\beta k} + T_{\beta k}) = 0. \tag{8}$$

### 4. The equation for the metric and the relativistic energy

The equation for the metric is obtained by equating to zero the sum of the terms containing variation $\delta g_{\alpha\beta}$ in (4):

$$-kR^{\alpha\beta} + \frac{k}{2} R g^{\alpha\beta} - k\Lambda g^{\alpha\beta} - \frac{1}{2c} s_\mu J^\mu g^{\alpha\beta} - \frac{1}{2c} T^{\alpha\beta} - \frac{1}{2c} A_\mu j^\mu g^{\alpha\beta} - \frac{1}{2c} W^{\alpha\beta} = 0. \tag{9}$$

Let us contract this equation by multiplying by the metric tensor, given that $g_{\alpha\beta} T^{\alpha\beta} = 0$, $g_{\alpha\beta} W^{\alpha\beta} = 0$, $g_{\alpha\beta} R^{\alpha\beta} = R$, $g_{\alpha\beta} g^{\alpha\beta} = 4$:

$$kR - 4k\Lambda - \frac{2}{c} s_\mu J^\mu - \frac{2}{c} A_\mu j^\mu = 0. \tag{10}$$

Substituting (10) in (9) simplifies the equation for the metric:

$$R^{\alpha\beta} - \frac{1}{4} R g^{\alpha\beta} = -\frac{1}{2ck} (T^{\alpha\beta} + W^{\alpha\beta}). \tag{11}$$

Solution of equation (11) was presented in [11] for the case of gravitational and electromagnetic fields, and the coefficient $k$ can be found by comparison with the experimental data.

Acting in the same way as in [18], we will write the gauge condition of the cosmological constant $\Lambda$. In view of (2) we obtain the following:

$$ck\Lambda = -s_\mu J^\mu - A_\mu j^\mu = -\left(D_\mu + u_\mu + \pi_\mu + \lambda_\mu + g_\mu + w_\mu\right)J^\mu - A_\mu j^\mu. \qquad (12)$$

In the complete physical system all the field sources in the form of four-currents are taken into account, and all the fields in such a system, acting on the matter, have the internal origin. According to (12), each matter unit has its own value $\Lambda$, depending on the four-potentials of the two general field components and the respective four-currents. We can assume that the initial state of the system under consideration was the motionless, scattered in space, low-density matter in the form of individual particles. Then under action of gravitation this matter was broken into fragments and was brought together in much denser system. In this case, the constant $\Lambda$ reflects the total energy density of the matter in its proper fields in the initial state.

From calibration of (12) from (10) the relation follows between the scalar curvature and the cosmological constant:

$$R = 2\Lambda. \qquad (13)$$

After applying the covariant derivative $\nabla_\beta$ to all the terms in (11), taking into account (8) and the equation $\nabla_\beta \left( R^{\alpha\beta} - \frac{1}{2} R g^{\alpha\beta} \right) = 0$ as the property of the Einstein tensor, we obtain the equation $\frac{1}{4} \nabla_\beta \left( R g^{\alpha\beta} \right) = 0$ or the equivalent equation $\nabla_\beta R = 0$. From (13) the equation $\nabla_\beta \Lambda = 0$ follows, and in view of (12) this leads to the equation that must hold inside the matter:

$$\nabla_\beta \left( s_\mu J^\mu + A_\mu j^\mu \right) = \nabla_\beta \left( D_\mu J^\mu + u_\mu J^\mu + \pi_\mu J^\mu + \lambda_\mu J^\mu + g_\mu J^\mu + w_\mu J^\mu + A_\mu j^\mu \right) = 0.$$

The energy of the system, consisting of the matter and the fields, can be calculated in the same way as in [18]. In view of the calibration of (12), for the energy we obtain the following:

$$E = \int \left( s_0 J^0 + A_0 j^0 + \frac{c^2}{16\pi\varpi} s_{\mu\nu} s^{\mu\nu} + \frac{c^2 \varepsilon_0}{4} F_{\mu\nu} F^{\mu\nu} \right) \sqrt{-g}\, dx^1 dx^2 dx^3.$$

The energy depends only on the time components of the four-potentials and four-currents and does not depend on their space components. For the 4-momentum of the system we obtain: $p^\mu = \left( \frac{E}{c}, \mathbf{p} \right) = \left( \frac{E}{c}, \frac{E}{c^2} \mathbf{v} \right)$, where $\mathbf{p}$ and $\mathbf{v}$ denote the momentum of the system and the velocity of the system's center of mass.

## 5. Conclusions

We will divide all the known fields into two groups. The first group includes the gravitational field, acceleration field, pressure field, dissipation field, strong interaction field, weak interaction field, and other vector fields, the source of which is the mass four-current $J^\mu$. This group of fields represents the mass component of the general field. The second (charge) component of the general field is the electromagnetic field, the source of which is the charge four-current $j^\mu$. All the field equations, the stress-energy tensors, the equation of the matter motion, the equation for the metric and the relativistic energy are found from the principle of least action in the covariant form. As it was shown in [21, 22], the potentials and strengths of all the fields obey the wave equation, and therefore in some cases they have the same dependence of the coordinates and time. In this case, the field potentials, field tensors and stress-energy tensors can be found using the standard procedure [23].

Both components of the general field are related to each other not only in the equations, but also in actual processes. Thus, the charge four-current is always accompanied by the corresponding mass four-current of the charge carriers, each of which has its own mass and velocity. In turn, the gravitational field, pressure field and other fields can generate some charge four-current. An example is the emergence of magnetic fields in cosmic objects in the electrokinetic model [24], due to the mechanism of charge separation and their simultaneous rotation with the matter of the objects.

Dividing the general field to two components is most naturally explained in the modernized Fatio-Le Sage's model [25, 26], where the vacuum field also has two components – the graviton field and the field of charged particles (praons). At the same time, one physical mechanism can be responsible both for the emergence of gravitational force [27] by means of the graviton field's action and for the emergence of electromagnetic interaction [28, 29] under the action of the field of charged particles.

In this model, the fluxes of the smallest particles of the vacuum field penetrate all bodies and perform the electromagnetic and gravitational interaction between the remote matter particles. The resulting different types of interaction between the matter particles can be represented as the action of the pressure field, acceleration field, dissipation field and other fields included in the general field. Thus the vacuum field components lead to the general field components.

The common mechanism of interaction of the vacuum and general fields helps to explain the reason for the fact that all the partial fields as the components of the general field can be described by the same Maxwell-type equations. For example, by solving the wave equations for the four-potentials of the acceleration field and the pressure field it is possible to find the temperature and the pressure inside the stars and planets [14] in good agreement with the calculations performed by other methods. This approach to the four partial fields has been successfully applied to estimation of the kinetic energy of the macroscopic system's particles performed by three different ways [15].